\documentclass[a4paper,11pt,twoside]{paper}
\usepackage[english]{babel}
\usepackage[utf8]{inputenc}
\usepackage{amsmath,amssymb,amsthm,mathrsfs,mathtools,graphicx,color}
\usepackage{hyperref}
\newtheoremstyle{sans}{\parskip}{\parskip}{\itshape}
                       {0pt}{\bfseries\sffamily}{.}{ }{}
\newtheoremstyle{sansplain}{\parskip}{\parskip}{}
                       {0pt}{\bfseries\sffamily}{.}{ }{}

\theoremstyle{sans}
\newtheorem{prop}{Proposition}[section]

\newtheorem{thm}[prop]{Theorem}
\newtheorem{lem}[prop]{Lemma}

\theoremstyle{sansplain}
\newtheorem{rem}[prop]{Remark}

\makeatletter
\@addtoreset{equation}{section}
\makeatother

\makeatletter
\@addtoreset{figure}{section}
\makeatother

\newcommand\C{\mathcal{C}}
\newcommand\Dc{\mathcal{D}}
\newcommand\Eb{\mathbb{E}}
\newcommand\Pb{\mathbb{P}}

\renewcommand{\geq}{\geqslant}

\def\DD{\displaystyle}

\newcommand\1{\leavevmode\hbox{\rm \small1\kern-0.35em\normalsize1}}
\newcommand\ind[1]{\1_{\{#1\}}}
\def\egaldef{\stackrel{\mbox{\tiny def}}{=}}

\newcommand{\dproof}{\noindent {\it Proof.} \quad}
\newcommand{\fproof}{\hfill $\blacksquare$ }
\begin{document}
\title{A Markovian Analysis of  IEEE~802.11 Broadcast Transmission Networks with Buffering}
\author{Guy Fayolle\thanks{INRIA  Paris-Rocquencourt, Domaine de Voluceau, BP 105, 78153 Le Chesnay Cedex, France. Email: {\tt Guy.Fayolle@inria.fr}}    \and
        Paul Muhlethaler \thanks{INRIA Paris-Rocquencourt, Domaine de Voluceau, BP 105, 78153 Le Chesnay Cedex, France. E.mail: {\tt paul.muhlethaler@inria.fr}}}
 \date{\today}
\maketitle
\begin{abstract}
The purpose of this paper is to  analyze the so-called back-off technique of the IEEE~802.11 protocol  in broadcast mode with waiting queues. In contrast to existing models, packets arriving  when a station (or node) is in back-off state are not discarded, but are stored in a buffer of infinite capacity. As in previous studies, the key point of our analysis hinges on the assumption that the time on the channel is viewed as a random succession of transmission slots (whose duration corresponds to the length of a packet) and mini-slots during which the back-off of the station is decremented. These events occur independently, with given probabilities. The state of a node is represented by a two-dimensional Markov chain in discrete-time, formed by the back-off counter and the number of packets at the station. Two models are proposed both of which  are shown to cope reasonably well with the physical principles of the protocol. The stabillity (ergodicity) conditions are obtained and interpreted in terms of maximum throughput. Several approximations related to these models are also discussed.
\end{abstract}
\smallskip

\emph{MSC $2010$ Subject Classification}: Primary 60J10; secondary 30D05, 30E99.

\section{Introduction}
Several studies~\cite{Bianchi,Wang} have recently been devoted  to the analysis of the IEEE~802.11 protocol,  both in the unicast and broadcast modes. Nonetheless, to the best of our knowledge, no model has taken into account the possibility of having  buffers to store arriving packets which, due to channel occupancy, can not be transmitted. When using the framework developed 
in~\cite{Bianchi,Wang}, it is impossible to rigorously evaluate the stability of the protocol. In this paper we use the same key assumption concerning  the slots as in~\cite{Bianchi,Wang}, but we couple it with a Markovian analysis of nodes using the IEEE~802.11 back-off with an infinite buffer for the packets.  We discuss how the way of sensing  slots on the channel  can give rise to two closely related models. The first one is analyzed in Section \ref{sec:greedy} and considers the so-called \emph{greedy mode}, while the second one in Section \ref{sec:fair} deals with a \emph{fair load} situation. In both cases, ergodicity conditions  and the value of the maximum throughput are obtained. The waiting time distribution of a packet is tackled  in Section \ref{sec:wtg}. 

\section{System parameters}\label{sec:para}
In contrast to the famous ALOHA protocol where the back-off does not take into account the activity of the channel, the back-off scheme of IEEE~802.11 monitors the channel in order to schedule the transmission of its pending packets. When a node receives a packet to send, it first selects a random  back-off number, which indicates the number of mini-slots the node has to wait before transmitting. A node with a pending packet senses the channel, and decreases its back-off counter by one each time an idle mini-slot is detected. When its back-off counter reaches the value zero, the node transmits the packet. 
 
 The following parameters are ubiquitous in the analytic studies of IEEE~802.11 broadcast: 
\begin{itemize}
\item $\sigma$, the duration of a mini-slot (MS), which is the time needed for a station to sense whether the channel is busy.
\item The backoff window size $W$, expressed as a multiple of $\sigma$.
\item $T$, the packet duration, a priori larger than $\sigma$. 
\end{itemize}
The three above parameters are really essential in the IEEE~802.11 back-off scheme. They lead to the main assumption that is usually made in the analysis of this scheme (see\cite{Bianchi,Wang}), stating that the channel consists in a random succession of \emph{full slots} (dedicated to packet transmission) and \emph{mini-slots} (short slots during which the channel is idle). To complete the description of the model, we also need to introduce the following three additional quantities. 
\begin{itemize}
\item $\lambda$, the Poisson  arrival rate at a station. 
\item $M$, the number of stations in the network.
\item $\tau$, the probability that a station will transmit a packet, stressing that this can take place only when its backoff counter has reached zero. We will return to this parameter later.
\end{itemize}

\section{Model of an isolated station in a greedy mode}\label{sec:greedy}
This section is devoted to the analysis of the basic component of the network, namely a single station operating in compliance with some of the principles of  the IEEE~801.11 protocol. 

The channel is supposed to be sampled at discrete time instants $Z_i$, so that it is sensed during consecutive time intervals $Z_i - Z_{i-1},i\ge1$, which are  equal  either to $T$ (a~\emph{normal slot}) or to $\sigma$ (a~\emph{mini-slot} (see Section \ref{sec:isol}). In some sense, we have thus introduced two embedded time-scales.

In general the selection between these two types of slots is random: a  slot is a normal slot with probability $r$, and a mini-slot with probability $1-r$. Nonetheless the node is said to be \emph{greedy}, since, when the back-off reaches 0, a transmission takes place in a slot which is then necessarily a normal slot of length $T$. In particular, this implies that a sequence of slots is not exactly obtained as  a pure outcome of repeated Bernoulli trials.

Now, starting from the basic one-dimensional models proposed in \cite{Bianchi,Wang}, we construct a two-dimensional Markov chain, taking into account the possibility of waiting queues at the stations.

An arbitrary station (or node)  will be represented by a stochastic process, which is a two-dimensional random walk  
$(K_{Z_i},N_{Z_i})_{i\ge1}$, where $Z_i$ is an embedded increasing sequence of discrete random times at which the chain is observed. Indeed, at time $Z_i$,  $K_{Z_i}$ stands for the value of the backoff counter and $N_{Z_i}$ is the number of packets in the waiting queue at a station. The Markovian evolution of $(K_{Z_i},N_{Z_i})$ is described in detail in the next section.
\subsection{Dynamics}\label{sec:isol}
Taking into account the definitions given in Section \ref{sec:para},  the sequence $Z_i,i\ge 1$ in the so-called \emph{greedy mode} satisfies the recursive stochastic relationship
\begin{equation}\label{eq:ZI1}
Z_{i+1} = Z_i + \Delta_i\bigl[\ind{K_{Z_i>0}} + \ind{K_{Z_i}=N_{Z_i}=0}\bigr] +
  T\ind{K_{Z_i}=0,N_{Z_i}>0},
\end{equation}
with
\begin{equation}\label{eq:delta}
\Delta_i = T\ind{B_i} + \sigma(1-\ind{B_i}),
\end{equation}
where $\ind{.}$ denotes the indicator function and $B_i$ is the event \emph{\{channel busy\}} at time $Z_i$. From the assumptions made above, $\{\ind{B_i},i\ge1\}$ form a sequence of independent identically distributed random varaibles with  $\Pb(B_i)=r$. 

Without entering a quite standard formalism, we just state that the underlying probability space for the random walk $(K_{Z_i},N_{Z_i})_{i\ge1}$ is obtained by combining the external Poisson arrival process and  the sequence of independent Bernoulli trials related to the $B_i$'s. 

Then we are in a position to write  Kolmogorov's forward equations for the Markov chain 
$(K_{Z_i},N_{Z_i})$.  The reader will see that the \emph{greediness}  appearing in the title of the section stems from the last term in the right-hand side member of~\eqref{eq:ZI1}.

Let us define the conditional probability $p(k,n;Z_i)\egaldef P(K_{Z_i}=k,N_{Z_i}=n;Z_i)$. Then, for all $k\ge0, n\ge1$, 
\begin{align}
p(k,n;Z_i+T)  &=\ind{k>0} \ind{B_i}\sum_{j=0}^{n }  \frac{e^{-\lambda T}(\lambda T)^j}{j!}p(k,n-j;Z_i) \nonumber\\[0.1cm]  
&+ \frac{1}{W+1}  \sum_{j=0}^{n }  \frac{e^{-\lambda T} (\lambda T)^j}{j!} p(0,n+1-j,Z_i)
\nonumber\\[0.1cm]
&+  \ind{B_i} \, \frac{1}{W+1}  \frac{e^{-\lambda T}(\lambda T)^n}{n!}  p(0,0;Z_i),
\label{eq:Kol1}\\[0.1cm]
p(k,n;Z_i+\sigma) & = \ind{k<W}(1-\ind{B_i}) \sum_{j=0}^{n } \frac{e^{-\lambda \sigma}   
(\lambda \sigma)^j}{j!}  p(k+1,n-j;Z_i) \nonumber\\[0.1cm]
&+(1-\ind{B_i}) \, \frac{1}{W+1} \frac{e^{-\lambda \sigma}(\lambda \sigma)^n}{n!}  p(0,0;Z_i). \label{eq:Kol2}
\end{align} 
These equations deserve some explanation.

\begin{itemize}
\item The first term of the right-hand member of~\eqref{eq:Kol1} corresponds to the transmission of a packet by another station when the current station is in back-off, noting that arrivals may occur during this slot. The second term corresponds to a packet transmission by the current station and the choice of a back-off number for the next pending packet; new arrivals may still occur. The last term concerns arrivals at an empty current station during the transmission of a packet by the network; a random back-off is selected for the first witing packet (if any). 
\item It is worth remarking that for $k=0$ the first term in the right-hand member of~\eqref{eq:Kol1} does not exist, since in this case it is impossible to reach the state $(0,n,Z_i)$ from any state $(0,j,Z_{i-1}), j\le n$. Also, the second term in~\eqref{eq:Kol1} indicates that a transmission takes place when the backoff counter equals  $0$. 
\item In~\eqref{eq:Kol2}, the first term of  the right-hand member corresponds to the current station decrementing its back-off counter; new arrivals may occur during the mini-slot, while the second term  takes into account  the new arrival during a mini-slot in the current station which is idle; if at least one packet has arrived a random back-off number is selected for this packet.
\end{itemize}
On the other hand, the following boundary equations hold for $k=0$ and for all $Z_i, i\ge1$.
 \begin{equation}\label{eq:bound1}
\!\! \begin{cases}
p(k,0;Z_i) =0,  \   \forall k\ge1,  \\[0.1cm]
p(0,0;Z_i+T) =  e^{-\lambda T}\bigl[p(0,1;Z_i) + \ind{B_i} \,p(0,0;Z_i)\bigr],\\[0.1cm]
p(0,0;Z_i+\sigma) =e^{-\lambda \sigma} (1-\ind{B_i})\,p(0,0;Z_i),
\end{cases}
\end{equation} 
where $p(0,0,Z_i)$ is the empty (or idle) state, through which the system exhibits some regeneration properties.

In the following, we focus on the steady state behaviour of  system~\eqref{eq:Kol1}-\eqref{eq:Kol2}-\eqref{eq:bound1}, which is simply obtained by letting  $i\to\infty$, since in this case  $Z_i\to\infty$ almost surely (and even surely\,!).

Setting 
\[
 p(k,n) \egaldef \lim_{i\to\infty} \Eb [p(k,n;Z_i)],
\]
 we obtain, for all $n\ge1, 0 \le k \le W$,
 \begin{align}\label{eq:Kol3}
p(k,n) & = \ind{k<W} (1-r)  \sum_{j=0}^{n } \frac{e^{-\lambda \sigma}  (\lambda \sigma)^j}{j!}  p(k+1,n-j) \nonumber\\[0.1cm]
&+\ind{k>0} \, r \sum_{j=0}^{n }  \frac{e^{-\lambda T}(\lambda T)^j}{j!}p(k,n-j) \nonumber\\[0.1cm]
&+ \frac{1}{W+1} \sum_{j=0}^{n }  \frac{e^{-\lambda T} (\lambda T)^j}{j!} p(0,n+1-j) 
\nonumber\\[0.1cm]
&+ \frac{1}{W+1} \Bigl[\frac{(1-r)e^{-\lambda \sigma}(\lambda \sigma)^n}{n!} 
+ \frac{r e^{-\lambda T}(\lambda T)^n}{n!} \Bigr] p(0,0),
\end{align} 
 and
  \begin{equation} \label{eq:bound2}
\begin{cases}
p(k,0) = 0,  \   \forall 1\le k \le W, \\[0.2cm]
 e^{-\lambda T}p(0,1) = p(0,0) \bigl[ 1-r e^{-\lambda T} - (1-r)  e^{-\lambda \sigma} \bigr].
\end{cases}
 \end{equation}
Let us observe  that here $r$ is an exogenous parameter, but later  in Sections \ref{sec:net1}, \ref{sec:net2} it will depend on the state of the system.

\subsection{Formal solution of the steady-state equations}\label{sec:formal1}
Let us introduce the generating functions
\[
F_k(x) = \sum_{n\ge0}p(k,n)x^n, \quad \forall 0\le k\le W,
\]
which will be sought to be \emph{analytic} in the unit disc 
\mbox{$\Dc=\{ x:|x|<1\}$} and continuous on the boundary $|x|=1$. 

\begin{lem} \label{lem:sys}
The generating functions $F_k(x), k\ge0$, satisfy the linear system 
\begin{equation}\label{eq:sys}
\begin{cases}
a(x) F_1(x) = \widetilde{F_0}(x) +C(x), \\[0.1cm]
a(x) F_{k+1}(x) = b(x) F_k(x) + C(x), \quad \forall 1\le k\le W-1, \\[0.1cm]
b(x) F_W(x) + C(x) = 0,
\end{cases}
\end{equation}
where
\begin{equation}\label{eq:def1}
\begin{cases}
\widetilde{F_0}(x) = F_0(x) -p(0,0) \\[0.1cm]
\DD a(x)= (1-r)e^{-\lambda\sigma(1-x)},\\[0.1cm]
\DD b(x) = 1-re^{-\lambda T(1-x)},
\end{cases}
\end{equation}
and 
\begin{equation}\label{eq:def2}
C(x) = \frac{-e^{-\lambda T(1-x)}}{(W+1)x} F_0(x) +\frac{p(0,0)}{W+1}\left[1+ 
\Bigl(\frac{1}{x}-r\Bigr)e^{-\lambda T(1-x)} -(1-r)e^{-\lambda\sigma(1-x)}\right].
\end{equation}
\end{lem}

\dproof
System~\eqref{eq:sys} is easily obtained after mutiplying~\eqref{eq:Kol3} by $x^n$, summing from $n\ge1$ onwards, and taking into account the first boundary condition in~\eqref{eq:bound2}. The final form of $C(x)$ in \eqref{eq:def2} is derived from the second boundary condition in~\eqref{eq:bound2}. More detail is given in Appendix~\ref{sec:app1}.
\fproof
\medskip
\begin{prop}\label{prop:F0}
Letting \[
u(x)\egaldef \frac{b(x)}{a(x)},
\] 
the functions $F_k(x), 1\le k \le W$, are given by the relations
\begin{equation}\label{eq:Fk}
F_k(x) = \frac{\widetilde{F_0}(x)}{a(x)}\left[\frac{u^{k-1}(x) - u^W(x)}{1-u^{W+1}(x)}\right],
\end{equation}
and $F_0(x)$ has the explicit form
\begin{equation}\label{eq:F0}
F_0(x) = p(0,0) \frac{Q(x)}{R(x)},
\end{equation}
where 
\begin{align}
 R(x)  & = \frac{e^{-\lambda T(1-x)}}{x} - \frac{(W+1)u^W(x)(1-u(x))}{1-u^{W+1}(x)}, \label{eq:R} \\[0.2cm]
 Q(x)  & =  \left[1+ \Bigl(\frac{1}{x}-r\Bigr)e^{-\lambda T(1-x)} -(1-r)e^{-\lambda\sigma(1-x)}\right]
  -  \frac{(W+1)u^W(x)(1-u(x))}{1-u^{W+1}(x)}.\label{eq:Q}
\end{align}
\end{prop}
\dproof
The result is  obtained in two steps. First a straightforward calculus exploiting  system~\eqref{eq:sys} yields~\eqref{eq:Fk}. Then, instantiating $k=W$ in equation~\eqref{eq:Fk} and comparing it with the third relation in~\eqref{eq:sys},
 we get
 \[
 u^W(x) \widetilde{F_0}(x) +C(x) \, \frac{1-u^{W+1}(x)}{1-u(x)}=0,
 \]
 which, by using \eqref{eq:def1},  leads to the final equations  \eqref{eq:F0},\eqref{eq:R},\eqref{eq:Q}. 
  \hfil \fproof
  
 The only remaining unknown $p(0,0)$ will now be obtained from the normalization condition  
\[
\sum_{k\ge0,n\ge0} p(k,n) =1.
\]
\begin{prop}\label{prop:p0}
Under ergodicity conditions, the probability $p(0,0)$ that the system is idle is given by the formula
\begin{equation}\label{eq:p00}
p(0,0) = \frac{1-\lambda B}{1- \lambda A +  \frac{\lambda W(B-A)}{2(1-r)}}
\end{equation}
where
\begin{equation} \label{eq:p01}
\begin{cases}
\DD A  & =  (1-r)(T-\sigma) + \frac{W[rT+(1-r)\sigma]}{2(1-r)}, \\[0.4cm]
\DD B & =  T + \frac{W[rT+(1-r)\sigma]}{2(1-r)}. 
\end{cases}
\end{equation}
\end{prop}
\dproof
By \eqref{eq:Fk}, since $u(1)=1$ and $a(1)=1-r$, we get 
\[
 F_0(1) + \frac{\widetilde{F_0}(1)}{1-r}\,\sum_{k=1}^W \frac{W-k+1}{W+1} =1,
\]
that is
\begin{equation}\label{eq:norm}
F_0(1) + \frac{W\widetilde{F_0}(1)}{2(1-r)} = 1.
\end{equation}
Hence, instantiating $x=1$ in equations \eqref{eq:F0},\eqref{eq:R}-\eqref{eq:Q}, using the relation 
\[
\lim_{x\to1} \frac{Q(x)}{R(x)} = \frac{Q'(1)}{R'(1)},
\]
since $Q(1)=R(1)=0$, where $Q'(1)$ and $R'(1)$ are the respective derivatives of $Q(x)$ and $R(x)$ at $x=1$,  from equation \eqref{eq:norm} we obtain
\[
p(0,0)\left[ \biggl(1+\frac{W}{2(1-r)}\biggr)\frac{Q'(1)}{R'(1)} -\frac{W}{2(1-r)}\right] = 1.
\]
Setting for the present calculation 
\[
f(x) =  \frac{(W+1)u^W(x)(1-u(x))}{1-u^{W+1}(x)}=  \frac{(W+1)u^W(x)}{\DD\sum_{0\le i\le W}u^i(x)},
\]
we have
\[
f'(1) = \frac{Wu'(1) }{2}= -\frac{\lambda W[rT+(1-r)\sigma]}{2(1-r)}.
\]
Hence, by \eqref{eq:R} and \eqref{eq:Q},
\begin{eqnarray*}
R'(1) &= &-1+\lambda T - f'(1) \,  = \,  -1+\lambda T + \frac{\lambda W[r T + (1-r)\sigma]}{2(1-r)},\\[0.2cm]
Q'(1)&=& -1 +(1-r) \lambda(T-\sigma) -f'(1) \,  = \, -1 +(1-r) \lambda(T-\sigma) +
\frac{\lambda W[rT+(1-r)\sigma]}{2(1-r)},
\end{eqnarray*}
which by an elementary computation yields equation \eqref{eq:p00}.  \fproof

\medskip
\begin{rem} \label{rem:tau}
 As we shall see in  Theorem \ref{th:ergo}, the system is ergodic if and only if
\[
R'(1)<0,
\]
in which case the following inequalities hold
\[
\begin{cases}
\DD Q'(1)\le R'(1)<0, \\[0.2cm]
\DD\frac{Q'(1)}{R'(1)}\ge1.
 \end{cases}
\]
Two limit cases can also be checked. 
\begin{itemize}
\item $\lambda\to0\ \Rightarrow p(0,0) \to 1$.
\item It should be observed that  $Q'(1)=R'(1)$ if and only if $r=\sigma=0$, which corresponds to the trivial situation $p(0,0)=1$.
\end{itemize}
\end{rem} \hfill$\square$

We bear in mind that  functions $F_k(x), k\ge0$, are sought to be analytic in the unit disc. Hence ergodicity conditions will be obtained from \eqref{eq:F0} by studying the possible zeros of $R(x)$ in the closed unit disk $\Dc$, noting \emph{en passant} that $Q(x)$ and $R(x)$ have no common root  in $\Dc$, but at $x=1$. This is the objective of the next theorem. 

\subsection{Stability condition of  the greedy model} With the notation of Proposition \ref{prop:p0}, we have the following 
\begin{thm} \label{th:ergo}The system is ergodic if and only if $R'(1)=\lambda B -1<0$, that is
\begin{equation} \label{eq:ergo}
 \lambda  <  \frac{1}{T \left[ 1 + \frac{rW}{2(1-r)}\right]+ \frac{W\sigma }{2}}.
\end{equation}
In other words, the ergodicity region in the parameter space $\{\lambda,\sigma,T,r\}$ is delimited by the surface 
 $S(\lambda,\sigma,T)$ expressed by the equation
 \[
 \lambda \bigl[(Wr+2(1-r))T + W(1-r)\sigma\bigr] =  2(1-r).
 \]
\end{thm}
\dproof 
Let
 \[
 g(x) = \frac{e^{-\lambda T(1-x)}[1-u^{W+1}(x)]}{1-u(x)}, \quad h(x)=(W+1)xu^W(x),
 \]
one can see at once that, for all $x\ne1$, the zeros of $R(x)$ in $\Dc$ coincide with the roots of
\[
g(x)-h(x) = 0.
\]
 From \eqref{eq:def1}
 \[
 u(x) = \frac{e^{\lambda\sigma(1-x)}[1-re^{-\lambda T(1-x)}]}{(1-r)},
 \]
 so that, $\forall x\in\Dc$,
 \[
 |u(x)| = \frac{|e^{\lambda\sigma(1-x)}|}{1-r} |1-re^{-\lambda T(1-x)}| \ge |e^{\lambda\sigma(1-x)}|\ge1,
 \]
 which yields
 \[
\left|\frac{1-u^{W+1}(x)}{1-u(x)}\right| = |1+u(x)+\cdots+u^W(x)| \le (W+1)|u^W(x)|.
 \]
Therefore, on the unit circle $|x|=1,x\ne1$, we have  $|g(x)|<|h(x)|$. Since the functions $g(x)$ and $h(x)$ are analytic in $\Dc$, we will be in a position to apply the well-known Rouch\'e's theorem (see e.g. \cite{Tit}), after carrying out an adequate modification of the unit circle $\C$ around the point $x=1$. Since 
\[
g(x)-h(x)= \frac{x(1-u^{W+1}(x))}{1-u(x)} R(x),
\]
we have
 \[
 g(x)-h(x) = (W+1)R'(1)(x-1) + o((x-1)^2).
  \]
 Let us assume  $R'(1)<0$. We construct a contour   $\C_\varepsilon$  consisting essentially of  the unit circle $\C$ continuously distorted around $x=1$ by a small notch (an arc of a circle of radius 
 $\varepsilon$), keeping the point $1$ inside the domain $\Dc_\varepsilon $ bounded by 
 $\C_\varepsilon$.  Then, choosing   $\varepsilon$ sufficiently small, we have
 \[
 |g(x)|<|h(x)|, \quad \forall x \in\C_\varepsilon.
 \]
Consequently, by Rouché's theorem, functions $h(x)$ and $h(x)-g(x)$ have the same number of zeros inside  $\Dc_\varepsilon$, i.e. exactly one located at $x=1$, since $h(x)$ vanishes only at $x=0$. We conclude that $F_0(x)$ given by \eqref{eq:F0} is analytic in $\Dc$ and the system is ergodic.

Conversely, if $R'(1)>0$, we can construct a contour $\Dc_\varepsilon$ that does include the point $x=1$, and the same argument shows that $R(x)$ has a root different from $1$ inside the unit disk $\Dc$, and in this case $F_0(x)$ is not analytic in $\Dc$. \fproof

\medskip
We observe that, on the one hand, for $r$ tending towards 1, $\lambda$ tends towards $0$. The activity outside  the current station is prominent, which means that no bandwidth is left for the current station. On the other hand, if $r\to0$, then $\lambda$ can reach the value $\lambda_{\max}= \frac{1}{T+W\sigma/2}$. In this case 
the current station can consume all the bandwidth, the term $\frac{W\sigma}{2}$ being the overhead due to the back-off scheme. 

\subsection{Dynamics of a network of greedy broadcasting stations}  \label{sec:net1}
 Consider a network consisting of $M$ identical stations operating in a broadcast mode together with a tagged station. The key feature of the protocol IEEE~802.11 ---namely a transmission is permitted only when the backoff counter is at zero--- can be reflected by the simple equality 
 \begin{equation}\label{eq:tau}
 \tau = \sum_{n\ge1} p(0,n).
 \end{equation}
Hence, from the point of view of the tagged station, the $M$ stations represent the outside world and the event  \emph{\{Channel busy\}} takes place with probability
\begin{equation}\label{eq:trans}
r=1- (1-\tau)^M, \quad M\ge0.
\end{equation} 

The problem is now to recast the ergodicity condition  \eqref{eq:ergo} given in Theorem \ref{th:ergo} in the light of equations \eqref{eq:tau}, \eqref{eq:trans}. According to Proposition \ref{prop:p0} and Remark \ref{rem:tau}, $\tau$ defined by \eqref{eq:tau}  satisfies   
\[
\tau = F_0(1) - p(0,0) = p(0,0)\left(\frac{Q'(1)}{R'(1)} -1\right) =  
 p(0,0)\,\frac{\lambda(B-A)}{1-\lambda B}, 
\]
 that is, by \eqref{eq:p00} and \eqref{eq:p01},
 \begin{equation}\label{eq:tau3}
 \tau =\frac{ \lambda(B-A)}{1-\lambda A + \frac{\lambda W(B-A)}{2(1-r)}} =
 \frac{ \lambda[rT+(1-r)\sigma]}{1-\lambda A + \frac{\lambda W(B-A)}{2(1-r)}} =
  \frac{ \lambda[rT+(1-r)\sigma]}{1-\lambda T +\lambda[rT+(1-r)\sigma]}.
 \end{equation}
Observe at once that the denominator in the last equality of \eqref{eq:tau3} necessarily imposes 
\begin{equation}\label{eq:l1}
\lambda T < 1,
\end{equation}
with an obvious interpretation. Henceforth condition \eqref{eq:l1} will be assumed in the rest of this section.

So, \eqref{eq:trans} and \eqref{eq:tau3} yield the following fixed point equation in  $r$, written as
 \begin{equation}\label{eq:s}
(1-r)^{1/M} = \frac{1- \lambda T}{1-\lambda (1-r)(T-\sigma)}.
\end{equation}
On the other hand, setting 
\begin{equation}\label{eq:z}
z=(1-r)^{1/M},
\end{equation}
one can also view  \eqref{eq:s} as a polynomial equation in $z$, namely
 \begin{equation}\label{eq:pz}
P(z)\egaldef \lambda(T-\sigma)z^{M+1} - z + (1-\lambda T)=0.
\end{equation}
\begin{lem}
Under condition  \eqref{eq:l1},  equation \eqref{eq:pz} viewed as an equation in $z$ has only one root in the the real interval $[0,1]$.
\end{lem}
\dproof 
Under the assumption  $\lambda T<1$, we have
\[
P(0)>0, \quad P(1)<0,
\]
and two cases must be considered.
\begin{enumerate}
\item $T<\sigma$ (purely academic situation). Then, on the interval $[0,1]$, the derivative $P'(z)$ in \eqref{eq:z} is negative, and $P(z)$ vanishes exactly once.
\item $T>\sigma$ (the real world\,!). On the unit circle $|z|=1$, we have
\[
|z +1-\lambda T| \ge \left| |z| - (1-\lambda T)\right|\ge \lambda T\ge \left|\lambda(T-\sigma)z^{M+1}\right|= \lambda(T-\sigma),
\]
and the result follows by an immediate application of Rouché's theorem.
\end{enumerate}
The proof of the lemma is concluded. \fproof

\begin{figure}[htb]
\vspace{1cm}
\centering\includegraphics[width=0.8\linewidth]{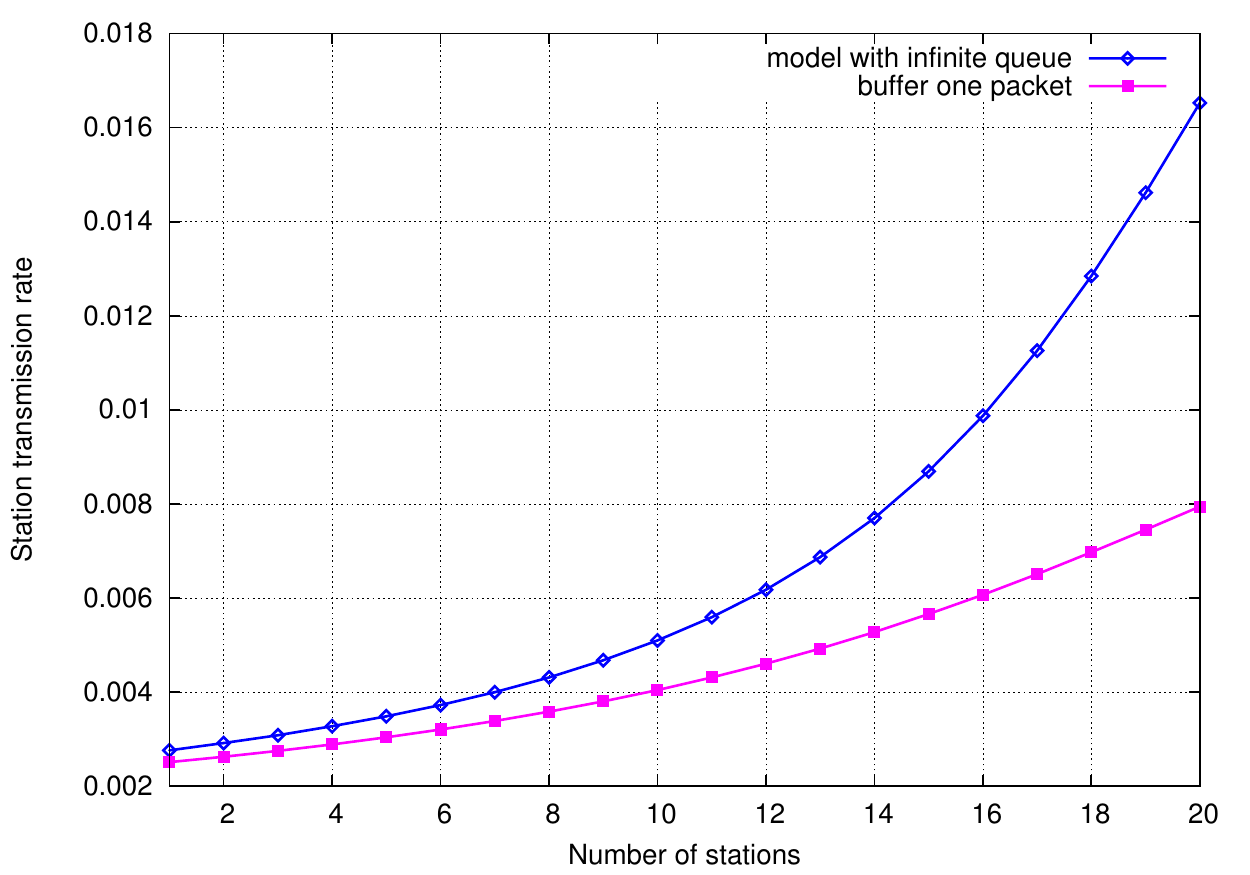} 
\caption{ \label{fig4.1} Computation of $\tau$ with and without any buffer, letting vary the number  of stations. }
\end{figure}

\begin{figure}[htb]
\vspace{1cm}
\centering\includegraphics[width=0.8\linewidth]{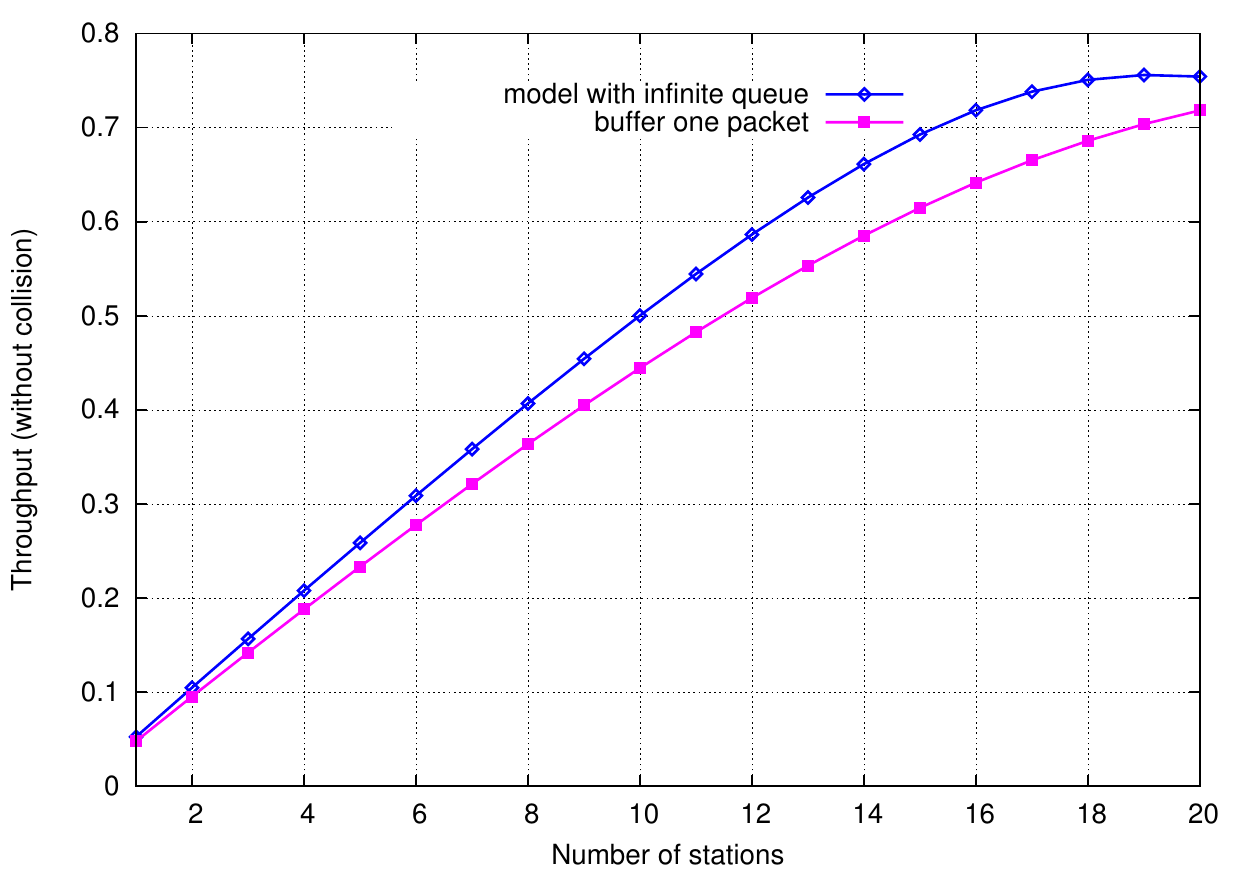} 
\caption{ \label{fig4.2} Network throughput without collision versus the number of stations. }
\end{figure}

\medskip
We adopt the following figures $\lambda=0.05$,  $T=1$, $\sigma=0.05$, $W=31$, and we vary $M$. We compute $\tau$ in our model and we compare it with the value of $\tau$ obtained in the model without any buffer of~\cite{Wang}.  In Figure~\ref{fig4.1}, we observe  that $\tau$ is larger in our model compared to the model without any buffer for the same input load. This is because the model without buffer drops packets when one packet arrives during the back-off period of a preceding packet. In Figure~\ref{fig4.2}, we compare the throughput without collision of our model with the throughput without collision of the model without buffer. For the same reason, the throughput without collision of our model is higher. 

 We are now in a position to compute the maximum throughput of the system under the stability condition \eqref{eq:ergo}. This is the subject of the following and final theorem.
 
 \begin{thm} \label{thm: GM} \mbox{ }
 \begin{enumerate}
  \item[1] The greedy broadcast network is ergodic if and only if
 \begin{equation}\label{eq:ergo2}
 2z^{M+1} > W(1-z),
 \end{equation}
 where $z$ defined by \eqref{eq:z} is  the unique real root  of  equation \eqref{eq:pz} in the interval $[0,1]$.
 \item[2]  When the system is ergodic, the maximum achievable throughput of the system takes one of the two equivalent forms
 \begin{equation}\label{eq:l2}
 \lambda_{max} = \frac{1-u}{T(1-u^{M+1}) + \sigma u^{M+1}} =  
  \frac{1-u}{ T + \frac{W(\sigma-T)(1-u)}{2}}, 
\end{equation}
where $u$ is the unique root on $[0,1]$ of the equation 
\begin{equation}\label{eq:W}
2 u^{M+1} =W(1- u).
\end{equation}
 \end{enumerate}
\end{thm}
\dproof
Rewrite $B$ given by \eqref{eq:p01} as
\[
B=T + \frac{W[(1-r)(\sigma-T) +T]}{2(1-r)},
\]
which, using \eqref{eq:pz}, yields the pleasant factorized form
\[
1-\lambda B = (1-\lambda T)\left[1-\frac{W(1-z)}{2z^{M+1}}\right].
\]
Hence, the necessary and sufficient ergodicity condition $\lambda B-1<0$ comes down to 
\[
1-\frac{W(1-z)}{2z^{M+1}} >0,
\]
which is exactly \eqref{eq:ergo2}. Finally, the quantity $\lambda_{\max}$ is simply obtained by saturating inequality \eqref{eq:ergo2}, which gives  \eqref{eq:W} and then \eqref{eq:l2}.  \fproof

\begin{figure}[htb]
\vspace{1cm}
\centering\includegraphics[width=0.8\linewidth]{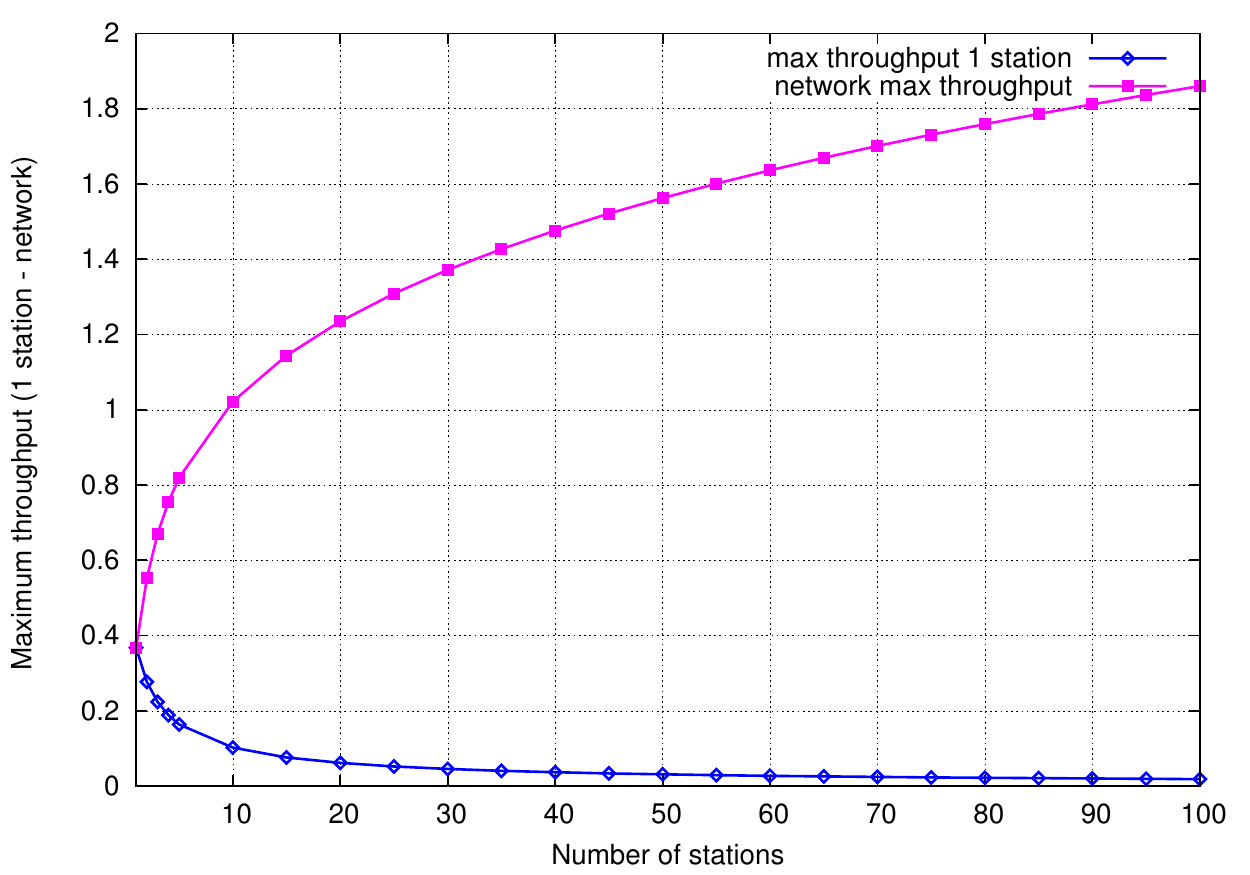} 
\caption{ \label{fig4.3} Network throughput versus the number of stations $M$.} 
\end{figure}

\begin{figure}[htb]
\vspace{1cm}
\centering\includegraphics[width=0.8\linewidth]{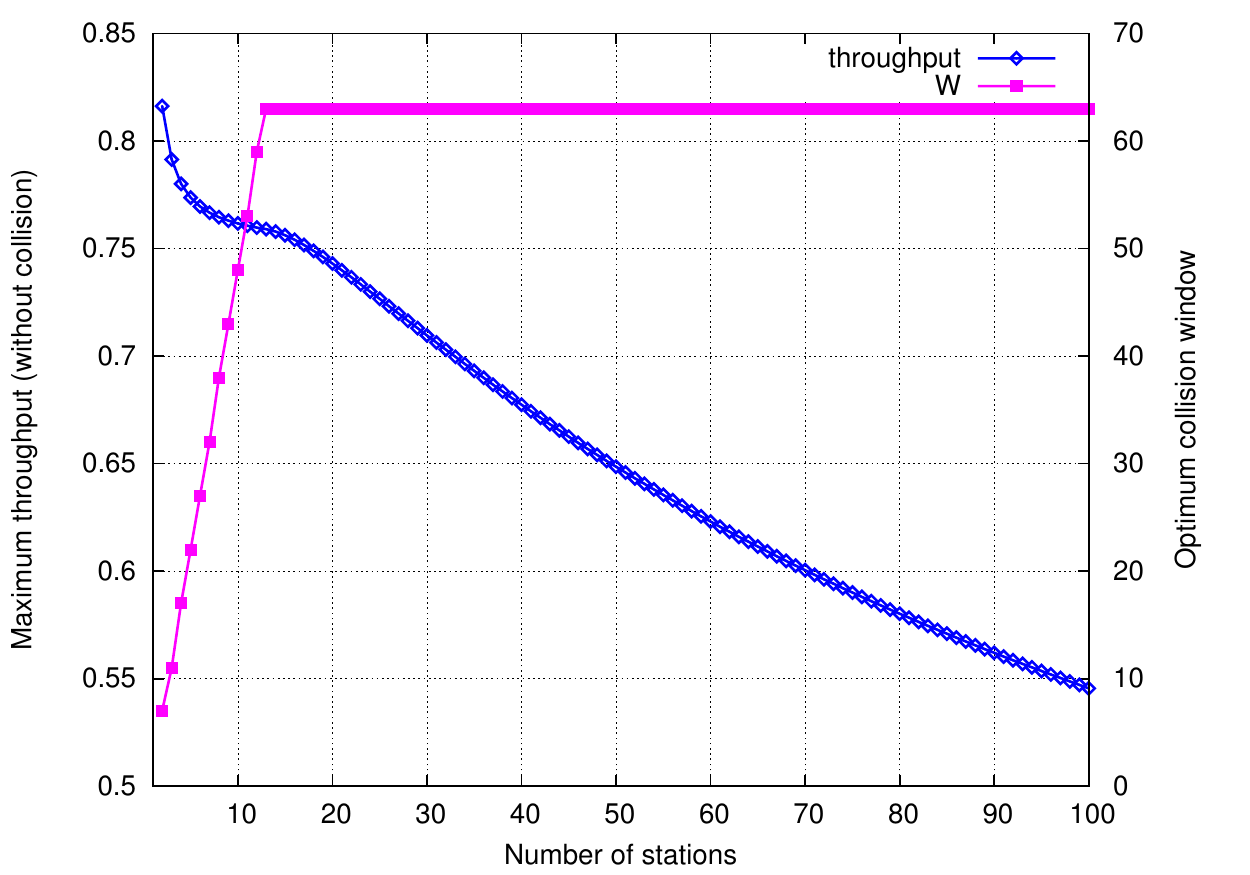} 
\caption{ \label{fig4.4} Optimized network throughput (without collision) and corresponding value of $W$ versus the number of stations $M$.}
\end{figure}

\bigskip 
We exploit formulas~\eqref{eq:l2} and~\eqref{eq:W}. We still have  $T=1$, $\sigma=0.05$, $W=31$, but 
$\lambda=0.05$, and we vary $M$ between $1$ and $100$. We plot the maximum throughput of one station and the cumulated throughput of all the stations which we call the network offered load. This is shown in Figure~\ref{fig4.3} where we vary the number $M$ of stations in the network. One can observe that the network offered load can be larger than $1$. This can be explained by the fact that two stations (or more) can send a packet during the same slot. The network offered load is the maximum total load  the network can submit while remaining stable. For a given number f stations, it is also possible to study the maximum network throughput (without collision) when we vary $W$. This study is presented in Figure~\ref{fig4.4}

\section{A fair load model} \label{sec:fair}
Here a station does not impose an extra load, as this was the case in the greedy model, and the sequence of times $Z_i,i\ge0$ is given by recursive scheme  
\[ 
Z_i = Z_{i-1} + \Delta_i,
\]
where $\Delta_i$ is still given by \eqref{eq:delta}. A station with a pending packet and a back-off counter at 0 may send its packet with probability $r$ or return to back-off with probability $1-r$. Thus, in contrast to Section~\ref{sec:greedy}, slots on the channel  are selected at random and form a renewal process:  a full slot (length $T$) with probability $r$, or a mini-slot (length $\sigma$) with probability $1-r$, independently of the state of the station. So, $r$ is considered as an exogenous parameter. But if the station is able to sense the channel, one could  think of $r$ being a function of the state, possibly estimated via some adaptive scheme like for example in~\cite{FGL} for Aloha-type systems. 
 
Mutatis mutandis with respect to Section \ref{sec:isol}, one  sees that the stationary distribution of a \emph{fair loaded}  station satisfies the following Kolmogorov's equations, for all $n\ge,0\le k\le W$,
 \begin{align}\label{eq:Kol4}
q(k,n) & = \ind{k<W} (1-r)  \sum_{i=0}^{n } \frac{e^{-\lambda \sigma} (\lambda \sigma)^i}{i!}  q(k+1,n-i) +\ind{k>0} \, r \sum_{i=0}^{n }  \frac{e^{-\lambda T}(\lambda T)^i}{i!}q(k,n-i) \nonumber\\[0.1cm]
&+ \frac{1}{W+1}\left[r \sum_{i=0}^{n }  \frac{e^{-\lambda T} (\lambda T)^i}{i!} q(0,n+1-i) 
+(1-r)  \sum_{i=0}^{n } \frac{e^{-\lambda \sigma} (\lambda \sigma)^i}{i!}  q(0,n-i)\right]
\nonumber\\[0.1cm]
&+ \frac{r e^{-\lambda T}(\lambda T)^n}{(W+1)n!}\,q(0,0),
\end{align} 
 and
  \begin{equation} \label{eq:bound3}
\begin{cases}
q(k,0) = 0,  \   \forall 1\le k \le W, \\[0.2cm]
 re^{-\lambda T}q(0,1) = q(0,0) \bigl[ 1-r e^{-\lambda T} - (1-r)  e^{-\lambda \sigma} \bigr].
\end{cases}
 \end{equation}

The first line of equation~\eqref{eq:Kol4} corresponds to the transition when the back-off counter is 
greater than or equal to $1$. If the current slot is a mini-slot (with probability $1-r$)  the back-off 
counter is decremented, otherwise if it is a normal slot (with probability $r$) and the back-off counter remains unchanged. In both cases new arrivals may occur and increment the number of pending packets.  The second line of  equation~\eqref{eq:Kol4} deals with the transition when the back-off counter is at $0$. With probability $r$, there is a transmission with a new draw of the back-off (if any pending packets) and new arrivals may occur. With probability $1-r$ the current slot is a mini-slot and the packet returns to back-off, and there still may be new arrivals. 
The third line of  equation~\eqref{eq:Kol4} corresponds to the transition from the idle state of the station  to a state where a back-off  value is selected with probability $1/(W+1)$.

Let us introduce as before the generating functions
\[ 
G_k(x) = \sum_{n\ge0}q(k,n)x^n, \quad 0 \le k\le W,
\]
 sought to be \emph{analytic} in the unit disc 
\mbox{$\Dc=\{ x:|x|<1\}$} and continuous on the boundary $|x|=1$. 
Then the following results hold, the explicit proofs of which are omitted, as they mimic the arguments of Section \ref{sec:formal1}. Some hints are given in Appendix \ref{sec:app2}.
 \begin{lem} \label{lem:sys2}
The generating functions $G_k(x), k\ge0$, satisfy the linear system 
\begin{equation} \label{eq:sys2}
\begin{cases}
a(x) G_1(x) = \widetilde{G_0}(x) +D(x), \\[0.1cm]
a(x) G_{k+1}(x) = b(x) G_k(x) + D(x), \quad \forall 1\le k\le W-1, \\[0.1cm]
b(x) G_W(x) + D(x) = 0,
\end{cases}
\end{equation}
where
\[ 
\widetilde{G_0}(x) = G_0(x)-q(0,0),
\]
$a(x),b(x)$ being given by \eqref{eq:def1}, and
\[ 
D(x) = \frac{-G_0(x)}{(W+1)}\left[\frac{re^{-\lambda T(1-x)}}{x} +(1-r)e^{-\lambda\sigma(1-x)}\right]
 + \frac{q(0,0)}{W+1} \left[ 1 + r\Bigl(\frac{1}{x}-1\Bigr)e^{-\lambda T(1-x)}\right].
\]
\end{lem}
 \begin{prop}\label{prop:G0}
Letting 
\[
u(x)\egaldef \frac{b(x)}{a(x)},
\] 
the functions $G_k(x), 1\le k \le W$, are given by the relations
\begin{equation}\label{eq:Gk}
G_k(x) = \frac{\widetilde{G_0}(x)}{a(x)}\left[\frac{u^{k-1}(x) - u^W(x)}{1-u^{W+1}(x)}\right],
\end{equation}
and $G_0(x)$ has the explicit form
\begin{equation}\label{eq:G0}
G_0(x) = q(0,0) \frac{\overline{Q}(x)}{\overline{R}(x)},
\end{equation}
where 
\begin{align}
 \overline{R}(x)  & = \frac{re^{-\lambda T(1-x)}}{x} +(1-r)e^{-\lambda\sigma(1-y)} 
 - \frac{(W+1)u^W(x)(1-u(x))}{1-u^{W+1}(x)}, \label{eq:Rbar} \\[0.2cm]
 \overline{Q}(x)  & =  \left[1+ r\Bigl(\frac{1}{x}-1\Bigr)e^{-\lambda T(1-x)}\right]
  -  \frac{(W+1)u^W(x)(1-u(x))}{1-u^{W+1}(x)}.\label{eq:Qbar}
\end{align}
\end{prop}
\dproof Along arguments quite similar to those used in Section \ref{sec:formal1}, one can show
\[
u^W(x) \widetilde{G_0}(x) +D(x) \, \frac{1-u^{W+1}(x)}{1-u(x)}=0,
\]
which directly yields  \eqref{eq:G0}, \eqref{eq:Rbar}, \eqref{eq:Qbar}. \fproof
\medskip 
\begin{thm} \label{th:ergo2} The fair load station is ergodic if and only if $\overline{R}'(1)<0$, that is
\begin{equation}\label{eq:ergo3}
 \lambda  <  \frac{r(1-r)}{\left[ 1-r +W/2 \right]\left[ rT +(1-r)\sigma\right]}.
\end{equation}
In this case, 
\begin{equation}\label{eq:q00}
q(0,0)= 1 - \frac{\lambda [rT+(1-r)\sigma] \left[1+ \frac{W}{2(1-r)}\right]}{r}.
\end{equation}
\end{thm}

\dproof As in Proposition \ref{prop:p0}, by using~\eqref{eq:Gk} and~\eqref{eq:G0}, the normalization condition becomes
 \[
 G_0(1) + \frac{W\widetilde{G_0}(1)}{2(1-r)} =1,
\]
whence, after some algebra
\begin{equation}\label{eq:q001}
q(0,0)= \frac{1}{\frac{W}{2(1-r)}\left(\frac{\overline{Q}'(1)}{\overline{R}'(1)}-1\right) + \frac{\overline{Q}'(1)}{\overline{R}'(1)}} = \frac{-\overline{R}'(1)}{r},
\end{equation}
 which is exactly equivalent to \eqref{eq:q00}. One can remark that when $r\searrow0$, the system remains ergodic if and only if 
\[
\frac{\lambda}{r} \nearrow \frac{1}{\sigma(1+W/2)},
\]
in which case $q(0,0)\searrow0$. \fproof

\subsection{Dynamics of a network of fair broadcasting stations} \label{sec:net2}
We proceed as in Section \ref{sec:net1}, setting 
 \begin{equation}\label{eq:rbar}
 r=1- (1-\overline{\tau})^M,
 \end{equation}
 with
  \begin{equation}\label{eq:taubar}
  \overline{\tau} = \sum_{n\ge1} q(0,n).
  \end{equation}
 Then the following result similar to Theorem \ref{thm: GM} holds.
  \begin{thm} \label{thm: FM} \mbox{ }
  The fair  broadcast network  is ergodic if and only if
 \[ 
 \lambda<\overline{\lambda}_{max},
 \]
 where the maximum achievable throughput $\overline{\lambda}_{max}$ takes the form 
 \begin{equation}\label{eq:l3}
\overline{\lambda}_{\max} =  \frac{1-u}{T + \frac{W\sigma(1-u)}{u(2+W)-W}}, 
\end{equation}
where $u$ is the unique root in $[0,1]$ of  \eqref{eq:W}, that is
\[
2 u^{M+1} =W(1- u).
\]
  \end{thm}
\dproof

We outline the arguments, as they mimic those of Theorem \ref{thm: GM}. The algebra is even simpler. Indeed, by~\eqref{eq:taubar} and~\eqref{eq:q001} one can write
\[
\overline{\tau} =  q(0,0)\left( \frac{\overline{Q}'(1)}{\overline{R}'(1)} -1\right)=\frac{\overline{R}'(1) -\overline{Q}'(1)}{r} = \frac{\lambda[rT+(1-r)\sigma]}{r},
\] 
so that, using~\eqref{eq:ergo3}
\begin{equation}\label{eq:taubar1}
 \lim_{\lambda\to\overline{\lambda}_{\max}}\overline{\tau}=  \frac{1}{1+\frac{W}{2(1-r)}}.
\end{equation}
Then, putting $1-r=u^M$, it appears that $u$ satisfies exactly \eqref{eq:W}, and we obtain
\[
\overline{\lambda}_{\max} = \frac{1}{\left[1+\frac{W}{2(1-r)}\right]\left[T+ \frac{(1-r)\sigma}{r}\right]} = 
\frac{1-u}{T+ \frac{(1-r)\sigma}{r}},
\]
which with \eqref{eq:rbar} and \eqref{eq:taubar1} yield \eqref{eq:l3}.
 \fproof

\begin{figure}[htb]
\vspace{1cm}
\centering\includegraphics[width=0.8\linewidth]{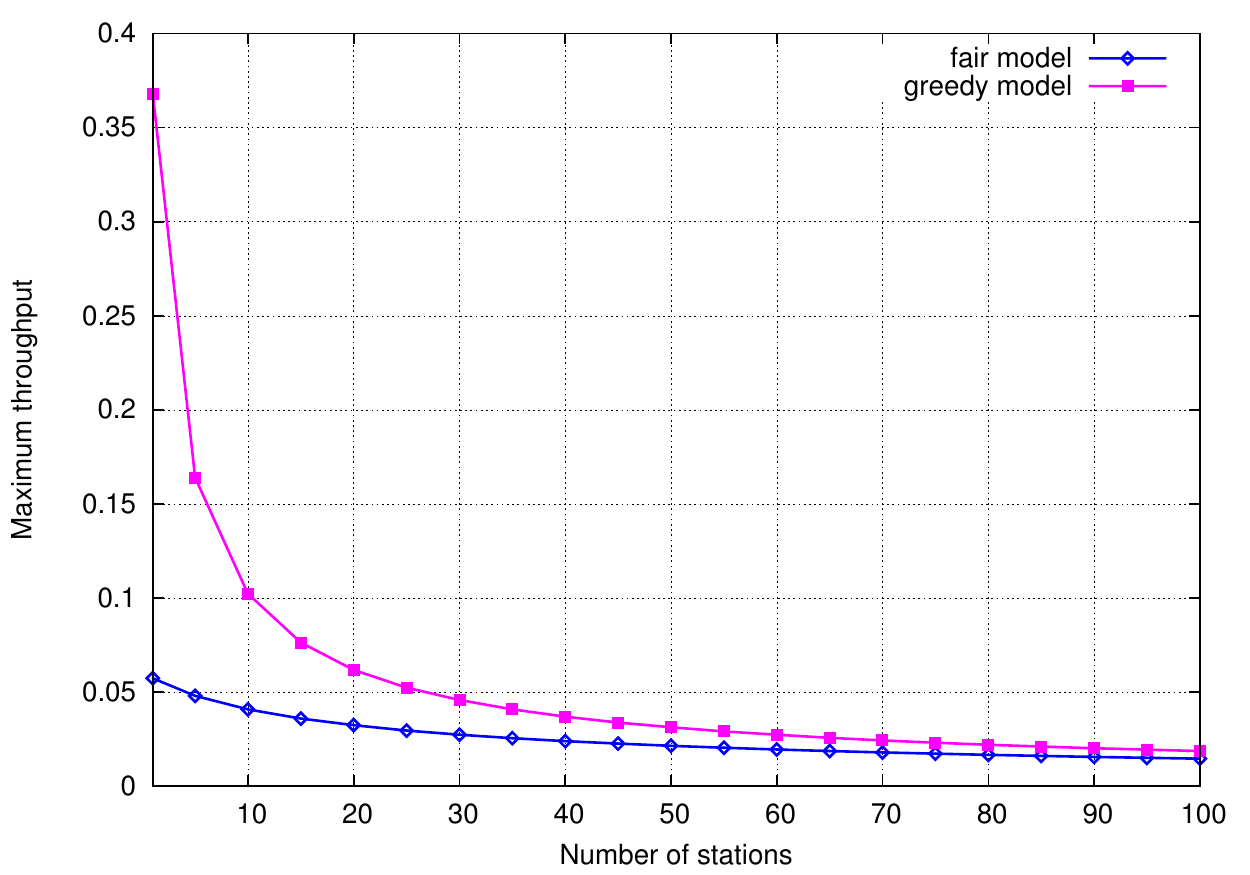} 
\caption{  \label{fig4.5} Maximum station throughput versus the number of stations $M$ in the fair and greedy models.}
\end{figure}

 \medskip
It is clearly worth comparing the two maxima $\lambda_{\max}$ and $\overline{\lambda}_{\max}$ given respectively by \eqref{eq:l2} and \eqref{eq:l3}. To this end, it suffices to check that the quantity
 \[
\delta = \frac{\sigma-T}{2} -  \frac{\sigma}{u(2+W)-W}= \frac{1}{2}\left[ \sigma-T - \frac{\sigma}{u(1-u)^M} \right], 
 \]
is clearly negative as $ u<1$.  Hence 
\[
\overline{\lambda}_{\max}<\lambda_{\max}.
\]  
We exploit the formula giving the maximum station throughput. We still have $\lambda=0.01$,  $T=1$, $\sigma=0.05$, $W=31$ and we vary $M$ between $1$ and $100$. We plot the maximum throughput of one station for the fair and greedy models. This is shown in Figure~\ref{fig4.5}
where we vary the number $M$ of stations in the network. As foreseen, the maximum achievable throughput is lower in the fair model compared to the greedy model.  However,  in both models $u$ is given by the unique equation \eqref{eq:W}, so that  the station transmission rates coincide, as well as the network throughputs without collision. 

\section{About the waiting time distribution in the greedy model}\label{sec:wtg}
From now on, the system will be assumed to evolve at \emph{steady state}. 
We intend to compute the stationary distribution of the virtual waiting time of a packet arriving at an arbitrary epoch $Z_i$. 

For an arbitrary probability distribution $F(.)$, denote by $F^*$ its Laplace-Stieltjes transform (LST)
\[F^*(s)\egaldef \int_0^{\infty} e^{-st}dF(t)dt, \quad \Re (s) \geq 0.
\]
Setting
\begin{equation}\label{eq:f}
f(s) =  \frac{(1-r)e^{-s\sigma}}{1-re^{-sT}}, \quad v(s) = \frac{e^{-sT}[f^{W+1}(s)-1]}{(W+1) (f(s)-1)},
\end{equation}
we have the following result.
 \begin{thm}\label{th:WT}
The Laplace transform of the stationary virtual waiting time distribution of a packet arriving at  time 
$Z_i$ has the form
\begin{equation} \label{eq:w1}
\psi^*(s) = \sum_{k=0}^W f^k(s) F_k(v(s)),
\end{equation}
In particular, the expectation $-\psi^*(0)'$ can be computed from formulas \eqref{eq:Fk},~\eqref{eq:F0}.
\end{thm}
\dproof

Let $A(k,n;Z_i)$ be the conditional waiting time of a packet  which at time $Z_i$ sees the system in the state $(K_{Z_i},N_{Z_i})=(k,n)$. With the notation of Section \ref{sec:isol}, we can write
\begin{equation}\label{eq:A}
A(k,n;Z_{i-1}) = (1-\ind{B_{i-1}}) [\sigma + A(k-1,n;Z_i)] + \ind{B_{i-1}}[T + A(k,n;Z_i)], \quad \forall k\ge1,
\end{equation}
remembering the transmission discipline at the buffer is  First-In-First-Out (FIFO), without forgetting  the boundary equation for $k=0$, namely
\begin{equation}\label{eq:Ab}
A(0,n;Z_{i-1}) = T+ \sum_{\ell=0}^W\ind{U=\ell} A(\ell,n;Z_i)], \quad \forall k\ge1,
\end{equation}
where $U$ is random variable uniformly distributed on the integers $0,1,\dots,W$.

Let $\varphi_i^*(s;k,n)$ be the LST corresponding to the distribution of  $A(k,n;Z_i)$. Then \eqref{eq:A} implies
 \[
 \varphi^*_{i-1}(s;k,n) = (1-r)e^{-s\sigma} \varphi^*_i(s;k-1,n) + re^{-sT} \varphi^*_i(s;k,n), \quad 
 \forall k\ge1,
 \]
 whence, letting $i\to\infty$ and $\varphi^* \egaldef \lim_{i\to\infty} \varphi^*_i$ and using the definition \eqref{eq:f},
 \begin{equation}\label{eq:wt2}
 \varphi^*(s;k,n) = f^k(s) \varphi^*(s;0,n).
  \end{equation}
Similarly, from \eqref{eq:Ab}, the LST $\varphi^*(s;0,n)$  satisfies the following  stationary recursive relationship  holds.
 \begin{equation}\label{eq:wt3}
\varphi^*(s;0,n) = \frac{e^{-sT}}{W+1} \sum_{\ell=0}^W \varphi^*(s;\ell,n-1) = 
\left[\frac{f^{W+1}(s)-1}{f(s)-1}\right]\frac{e^{-sT}\varphi^*(s;0,n-1)}{W+1}.
\end{equation}
Setting $\varphi^*(s;0,0)\equiv1$, we get from \eqref{eq:f}, \eqref{eq:wt2}, \eqref{eq:wt3},
 \[
\varphi^*(s;k,n) =  f^k(s)v(s)^n,
\]
and the sought waiting time distribution is given by
\[
\psi^*(s) = \sum_{k=0}^W\sum_{n=0}^\infty p(k,n) \varphi^*(s;k,n) 
\]
which reduces immediately  to \eqref{eq:w1}. \fproof 
\medskip
\begin{rem}
To compute  the waiting-time distribution of a packet at  arrival epoch, one has to switch from a discrete-time model to a continous-time one. We leave this question as an exercise, just quoting that it can be solved in  two steps.
\begin{itemize}
\item [1] First, find the distribution of the stationary Markov chain $(K_t,N_t)$  from $(K_{Z_i},N_{Z_i})$, for instance by applying a Palm probability inversion formula to the marked point process $Z_i$ with marks $(K_{Z_i},N_{Z_i})$). But a rather direct (although somehow tedious) is to proceed as in the derivation of residual life times for renewal processes: fix the unique interval 
$[Z_{i(\tau)},Z_{i(\tau)+1}]$ containing the arrival instant $\tau$, which allows to derive the number of new packets arrived between $Z_{i(\tau)}$ and $Z_{i(\tau)} +\tau$. Here the stationary sequence $Z_i$ satisfies
\[
\lim_{i\to\infty} \Eb (Z_{i+1}-Z_i) = [r+\alpha(1-r)]\,T + (1-r)(1-\alpha)\,\sigma,
\]
where $\alpha=F_0(1) - p(0,0)$.
\item[2] Then proceed as in the proof of  Theorem \ref{th:WT}.
\end{itemize}
\end{rem} \hfill$\square$

\section{Partial conclusion}
We have proposed two models for a CSMA protocol using the back-off of IEEE 802.11, one being
very close to the real  protocol for broadcast packets. In contrast to existing works, we assumed that nodes can store arriving packets. With a classical formalism of normal slot and mini-slots, it was possible to compute several important characteristic parameters:  station transmission rate, stability conditions, delay for a random packet. 

It is worth to emphasize that these protocols can stabilize \emph{without external control}, in contrast to many famous pioneers like ALOHA, CSMA, etc. In this respect, the window-backoff mechanism  exhibits some common features with the so-called tree (or stack) algorithms, which are stable provided that the external load remains below some critical value. 

The numerical results  obtained so far in this paper seem to match the performance of the IEEEE 802.11 broadcast protocol.  More simulations (left as future work) should very certainly confirm this matching. At last, studying  the IEEE 802.11 protocol for point-to-point traffic under the same hypothesis of infinite buffer, via the same mathematical approach, might be an interesting follow-up of this paper.
 
\section*{Acknowledgements} This paper is dedicated to Erol Gelenbe for his $70$th birthday. In particular, the first author remembers the old days (1975) of a cooperation on a model of the ALOHA protocol, a topic which initiated his indefectible interest in stochastic modelling. 
\bigskip
 \begin{center}
      {\bf APPENDIX}
    \end{center}
\appendix
\section{Equation  \eqref{eq:sys} of the greedy model}  \label{sec:app1}
Next is the intermediate step leading to system \eqref{eq:sys}. Indeed, mutiplying \eqref{eq:Kol3} by $x^n$, summing from $n\ge1$ onwards, and taking into account the first boundary condition in  \eqref{eq:bound2}, namely $p(k,0)\equiv0, \forall k\ge1$, we get
\begin{align*}
 F_k(x) & =  \ind{k<W}(1-r)e^{-\lambda\sigma (1-x)}F_{k+1}(x) +\ind{k>0} r e^{-\lambda T(1-x)}F_k(x) 
 \\[0.1cm]
 &+ \frac{1}{(W+1)x} \left[e^{-\lambda T(1-x)}(F_0(x) - p(0,0)) -p(0,1)e^{-\lambda T} x\right] 
 \\[0,1cm]
 &+ \frac{p(0,0)}{(W+1)}\left[r(e^{-\lambda T(1-x)} - e^{-\lambda T}) + (1-r)(e^{-\lambda\sigma(1-x)}-e^{-\lambda\sigma}) \right],
 \end{align*}
which yields in particular 
\[
a(x) F_{k+1}(x)= b(y)F_k(x) + C(x), \ \forall 1\le k \le W-1,
\]
whence \eqref{eq:sys} follows easily.
\section{Equation  \eqref{eq:sys2} of the fair load model model}  \label{sec:app2}
Just as in the case of the greedy model, \eqref{eq:sys2} is obtained by means of  the following system, derived from~\eqref{eq:Kol4},\eqref{eq:bound3},
\begin{align*}
 G_k(x) & =  \ind{k=0}q(0,0) + \ind{k<W}(1-r)e^{-\lambda\sigma (1-x)}G_{k+1}(x) 
 +\ind{k>0} r e^{-\lambda T(1-x)}G_k(x) \\[0.1cm]
 &+ \frac{r}{(W+1)x} \left[e^{-\lambda T(1-x)}(G_0(x) - q(0,0)) -p(0,1)e^{-\lambda T} x\right] 
 \\[0,1cm]
&+ \frac{1-r}{(W+1)}\left[e^{-\lambda\sigma(1-x)}G_0(x)-q(0,0)e^{-\lambda\sigma}\right] 
 + \frac{rq(0,0)(e^{-\lambda T(1-x)} - e^{-\lambda T})}{W+1}.
 \end{align*}


\end{document}